\newcommand{\be}{\begin{eqnarray}}
\newcommand{\ee}{\end{eqnarray}}

\documentclass[twocolumn,aps,showpacs]{revtex4}
\usepackage{graphics}
\begin{document}
\title{Finite size effects on pion spectra
       in relativistic heavy-ion collisions}
\author{Alejandro Ayala$^\dagger$, Eleazar Cuautle$^\dagger$, J.
Magnin$^\ddagger$, Luis Manuel Monta\~no$^*$ and Alfredo Raya$^\dagger$}   
\affiliation{$^\dagger$Instituto de Ciencias Nucleares, Universidad
Nacional Aut\'onoma de M\'exico, Apartado Postal 70-543, M\'exico
Distrito Federal 04510, M\'exico.\\
$^\ddagger$Centro Brasileiro de Pesquisas Fisicas, Rua Dr. Xavier
Sigaud 150-Urca CEP 22290-180, Rio de Janeiro, Brazil.\\
$^*$Centro de Investigaci\'on y de Estudios Avanzados del IPN,
Apartado Postal 14-740,
M\'exico Distrito Federal 07000, M\'exico}

\begin{abstract}

We compute the pion inclusive transverse momentum distribution
assuming thermal equilibrium together with transverse flow and
accounting for finite size effects and energy loss at the time of
decoupling. We compare to data on mid-rapidity pions produced in
central collisions in RHIC at $\sqrt{s_{NN}}=200$ GeV. We find that a
finite size for the system of emitting particles results in a
power-like fall-off of the spectra that follows the data up to larger
$p_t$ values, as compared to a simple thermal model. 

\end{abstract}


\maketitle

\date{\today}

\section{Introduction}

In recent years, strong evidence has been found in favor of the
production of matter composed of the fundamental QCD degrees of
freedom in central Au+Au collisions at the relativistic heavy-ion
collider (RHIC)~\cite{Bellwied, STAR}. Some of the quantitative
understanding about the signals supporting these findings are
associated to intermediate or high $p_t$ phenomena in particle
spectra. Among these, the depletion in the production 
of large $p_t$ charged hadrons and neutral pions, as compared to
extrapolations of data on p+p collisions to RHIC
energies~\cite{quenching}, suggests 
that the fragmenting partons that give rise to final hadrons, loose
energy on their way through the dense partonic matter believed to be
formed during the early stages of the reaction. This phenomenon has
been named jet quenching~\cite{Gyulassy} and has received
stronger experimental support from the results of the d+Au mode at
RHIC~\cite{dAu}. Moreover, the behavior of the proton to pion ratio for
intermediate $p_t\sim 2$ GeV~\cite{PHENIXBM} suggests that an important
mechanism of hadron production at RHIC is the thermal recombination of
free quarks~\cite{recomb} during the evolution of the collision. 

In spite of the success of this conceptual framework to interpret RHIC
data, there is an ingredient commonly overlooked: the fact that all
these phenomena take place during small time scales and consequently
within small volumes.

Although not commonly considered, it is certainly true that small size
effects are important in the description of statistical systems. For
example, finite size effects have influence on the late-stage growth
of nucleated bubbles during a first order phase
transition~\cite{Raju}. Also, in the statistical hadronization model,
the states used for the description are in general different to the
observed asymptotic ones but such difference is not an issue unless
the volume of the system is small, typically of order
${\mathcal{O}}(10\ {\mbox fm})^3$~\cite{Becattini}. Finite size
effects are also known to influence the interpretation of the
correlation lengths in Hanbury Brown-Twiss analysis in the context of
relativistic heavy-ion collisions~\cite{Zhang, Ayalaint}.

Finite size effects on particle spectra have also been recently
investigated~\cite{Mostafa, Ayalanoex, Ayalaex}, where these were referred
to as {\it boundary effects}. The scenario considered in the last two of these
studies resorts to a mean field treatment of the 
interactions affecting the most abundantly produced particles in the
reactions, namely pions. It was argued that prior to kinetic
freeze-out, the pion mean free path is smaller than the pion
interaction range and therefore that during this stage, pions are
still strongly interacting, mainly attractively, among themselves. The 
system resembles more a liquid than a gas, with a surface
tension acting as a reflecting boundary. When assuming thermalization
at freeze-out, the pion spectrum can be computed from the sum of the
square of the wave function in momentum space weighed by the thermal
occupation number for each state. 

Qualitatively, the behavior of particle spectra including this finite
size effects deviates from a simple exponential fall-off at high 
momentum since from the Heisenberg uncertainty principle, the more
localized the states are in coordinate space, the wider their
spread will be in momentum space. In terms of the discrete set of energy
states describing the pion system, this behavior can be understood as
arising from a higher density of states at large energy as compared to
a calculation without boundary.

In this work we compute the transverse momentum distribution for pions
assuming thermal equilibrium together with transverse flow and
accounting for finite size effects at decoupling. By comparing to data
on pion spectra on Au+Au collisions at $\sqrt{s_{NN}}=200$
GeV~\cite{PHENIXBM, PHENIX2}, we show that for temperatures and
collective transverse flow within the commonly accepted values, the
transverse momentum distributions can be described up to larger values
of $p_t$ as compared to a simple exponential fall-off. 

The paper is organized as follows: In Sec.~\ref{II}, we present the
basics of the model to compute the pion spectra including finite size
effects. In Sec.~\ref{III}, we compute the thermal pion spectrum
including the effects of transverse flow together with the
description of finite size effects. We carry out a systematic study of
the behavior of the spectra when varying the parameters involved. In
Sec.~\ref{IV} we compare the calculation to data on pion transverse
distributions from Au+Au collisions at $\sqrt{s_{NN}}=200$ GeV and
show how this model pushes the thermal component of the spectrum to larger
$p_t$ values as compared to a simple exponential
description. Finally we summarize and conclude in Sec.~\ref{concl}.

%%%%%%%%%%%%%%%%%%%%%%%%%%%%%%%%%%%%%%%%%%%%%%%%%%%%%%%%
%\vspace{0.9cm}
\begin{figure}[t!] % fig1
%\vspace{0.4cm}
{\centering
\resizebox*{0.4\textwidth}
{0.34\textheight}{\includegraphics{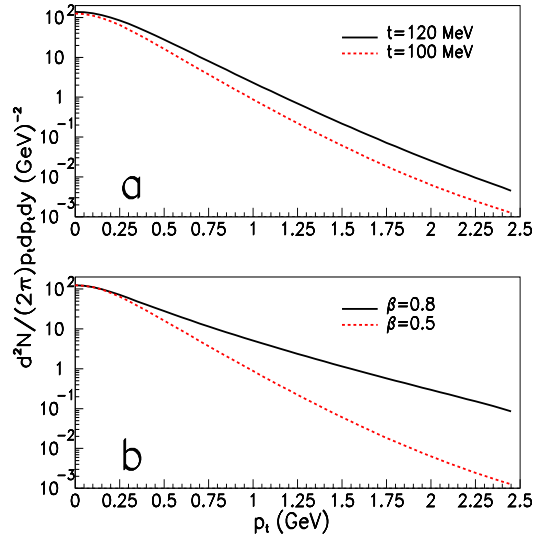}}
\par}
\caption{Invariant pion distribution as a function of $p_t$ for
different combinations of the parameters $T$ and $\beta$ for a fixed
value $R=8$ fm. The normalization is arbitrary.} 
\label{fig1}
\end{figure}
%%%%%%%%%%%%%%%%%%%%%%%%%%%%%%%%%%%%%%%%%%%%%%%%%%%%%%%%

\section{The model}\label{II} 

When the system of particles can be considered as confined and their wave
functions as satisfying a given condition at the boundary just before freeze 
out, the energy states form a discrete set. The shape of
the volume within the confining boundary is certainly an important
issue in case the model aims to describe transverse momentum particle
distribution at all rapidities. However, in this work we will restrict
ourselves to describe spectra at central rapidities. We thus consider
a scenario in which the system of particles of a given species is in
thermal equilibrium and is confined within a sphere 
of radius $R$ (fireball) as viewed from the center of mass of the colliding
nuclei at the time of decoupling. This time needs not be the same over
the entire reaction volume. Nevertheless, in the spirit of the
fireball model we consider that decoupling takes place over a constant
time surface in space-time. This assumption should be essentially
correct if the freeze-out interval is short compared to the system's
life time.

The solutions that incorporate the effects of a finite size system
have been found in Ref.~\cite{Ayalanoex}. They are given as the 
stationary solutions to the equation
\be
   \left(\frac{\partial^2}{\partial t^2} -
   \nabla^2 + m^2\right)
   \psi({\mathbf r},t)=0
   \label{eq:neweq}
\ee
subject to the condition
\be
   \psi(|{\mathbf r}|=R,t)=0\, ,
   \label{eq:cond}
\ee
and also finite at the origin. The normalized stationary states are
\be
   \psi_{nlm'}({\mathbf r},t)&=&\frac{1}{R\ J_{l+3/2}(k_{nl}R)}
   \left(\frac{1}{rE_{nl}}\right)^{1/2}\nonumber\\
   &&Y_{lm'}(\hat{r})J_{l+1/2}(k_{nl}r)e^{-iE_{nl}t}
   \label{eq:solnew}
\ee
where $J_\nu$ is a Bessel function of the first kind and $Y_{lm'}$ is a 
spherical harmonic. The parameters $k_{nl}$ are related to the energy  
eigenvalues $E_{nl}$ by
\be
   E_{nl}^2=k_{nl}^2+m^2
   \label{eq:param}
\ee
and are given as the solutions to
\be
   J_{l+1/2}(k_{nl}R)=0\, .
   \label{eq:encond}
\ee

%%%%%%%%%%%%%%%%%%%%%%%%%%%%%%%%%%%%%%%%%%%%%%%%%%%%%%%%
%\vspace{0.9cm}
\begin{figure}[t!] % fig2
%\vspace{0.4cm}
{\centering
\resizebox*{0.4\textwidth}
{0.34\textheight}{\includegraphics{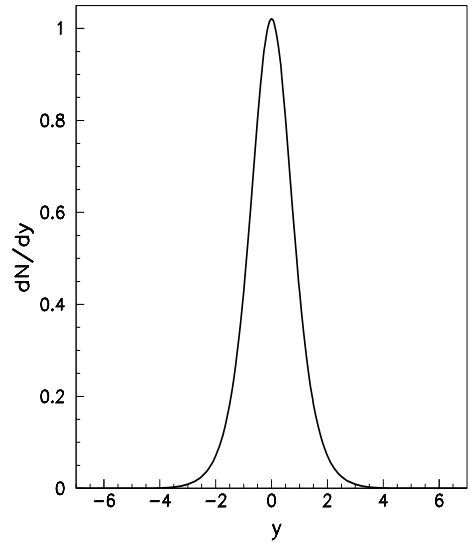}}
\par}
\caption{Pion rapidity distribution for $T=120$ MeV,
$\beta=0.6$ and $R=8$ fm. The normalization is arbitrary.} 
\label{fig2}
\end{figure}
%%%%%%%%%%%%%%%%%%%%%%%%%%%%%%%%%%%%%%%%%%%%%%%%%%%%%%%%

Thus, the eigenfunctions for the bound system of particles are given
explicitly by Eq.~(\ref{eq:solnew}). These 
are analytical, albeit given in terms of Bessel
functions. 

\section{Thermal spectrum}\label{III}
 
The contribution to the thermal particle invariant distribution from a
state with quantum numbers $\{n,l,m'\}$ is given by
\be
   E\frac{d^3N_{nlm'}}{d^3p}=
   \int \frac{d\sigma}{(2\pi)^3}(k_{nl}\cdot
   u) f(k_{nl}\cdot v){\mathcal{W}}_{nlm'}({\mathbf{p}},{\mathbf{r}}),
   \label{distone}
\ee
where ${\mathcal{W}}_{nlm'}({\mathbf{p}},{\mathbf{r}})$ is the Wigner
transform and $f(k_{nl}\cdot v)$ the thermal occupation factor of the
state $\psi_{nlm'}$, respectively. The four-vectors $v^\mu$ and $u^\mu$ 
represent the collective flow four-velocity and a four-vector of
magnitude one, normal to the freeze-out hypersurface $\sigma$,
respectively.

%%%%%%%%%%%%%%%%%%%%%%%%%%%%%%%%%%%%%%%%%%%%%%%%%%%%%%%%
%\vspace{0.9cm}
\begin{figure}[t!] % fig3
%\vspace{0.4cm}
{\centering
\resizebox*{0.4\textwidth}
{0.3\textheight}{\includegraphics{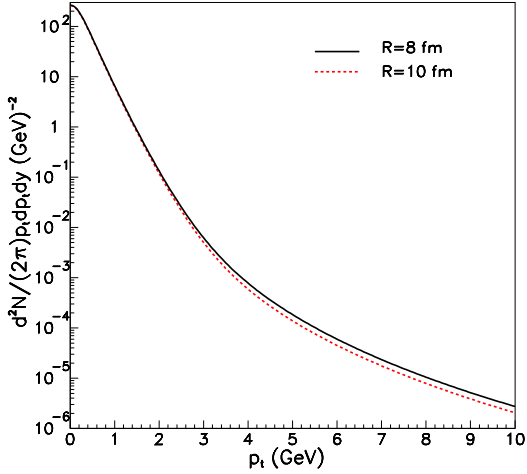}}
\par}
\caption{Invariant pion distribution as a function of $p_t$ for
two values of the parameter $R$ and fixed
values $T=120$ MeV and $\beta=0.6$. The normalization is arbitrary.} 
\label{fig3}
\end{figure}
%%%%%%%%%%%%%%%%%%%%%%%%%%%%%%%%%%%%%%%%%%%%%%%%%%%%%%%%

For the system of bound bosons, the Wigner transform of a given state
is defined as
\be
   {\mathcal{W}}_{nlm'}({\mathbf{p}},{\mathbf{r}})&=&
   \int d^3r' e^{-i{\mathbf{p}}\cdot{\mathbf{r}}}\nonumber\\
   &&\psi^*_{nlm'}({\mathbf{r}}+\frac{\mathbf{r'}}{2},t)
   \frac{\stackrel{\leftrightarrow}{\partial}}{\partial t}
   \psi_{nlm'}({\mathbf{r}}-\frac{\mathbf{r'}}{2},t).
   \label{Wigtra}
\ee
In order to consider a situation where freeze-out happens at a fixed
time and within a spherical volume of radius $R$, the unit
four-vector $u^\mu$ can be chosen as $u^\mu=(1,{\mathbf 0})$. 

To keep matters simple, we consider a thermal occupation factor of
the Maxwell-Boltzmann kind, namely
\be
   f(k_{nl}\cdot v)=e^{-k_{nl}\cdot v/T},
   \label{occupationfac}
\ee 
where $T$ is the system's temperature. The four-vector $v^\mu$ is
parametrized as
\be
   v^\mu=\gamma(1,{\mathbf v}),
   \label{fourvel}
\ee
and we choose a radial profile for the vector ${\mathbf v}$ such as
\be
   {\mathbf v}=\beta\frac{\mathbf r}{R},
   \label{tresvel}
\ee
where the parameter $\beta$ represents the surface expansion
velocity. Notice that, the radial profile for the expansion velocity
considered in Eq.~(\ref{tresvel}) can be taken as the transverse component of
the flow only for the description of central rapidity data. Correspondingly,
the gamma factor is given by  
\be
   \gamma=\frac{1}{\sqrt{1-\left(\beta\frac{r}{R}\right)^2}}.
   \label{gammafac}
\ee
Nonetheless, in order to continue to keep matters as simple as
possible, and be able to analytically perform the integrations in
Eq.~(\ref{distone}), we will instead consider that the gamma factor is
a constant evaluated at the average transverse expansion velocity, namely,
\be
   \gamma\rightarrow\bar{\gamma}=
   \frac{1}{\sqrt{1-(3\beta /4)^2}}.
   \label{gammaaver}
\ee
We take the four-vector $k_{nl}^\mu$ given by
\be
   k_{nl}^\mu=(E_{nl},{\mathbf k}_{nl}),
   \label{fourk}
\ee
and choose ${\mathbf k}_{nl}\parallel{\mathbf p}$. This choice is
motivated from the continuum, boundless limit, where the
relativistically invariant exponent in the thermal occupation factor
becomes $\gamma(E-{\mathbf p}\cdot{\mathbf v})$.

%%%%%%%%%%%%%%%%%%%%%%%%%%%%%%%%%%%%%%%%%%%%%%%%%%%%%%%%
%\vspace{0.9cm}
\begin{figure}[t!] % fig4
%\vspace{0.4cm}
{\centering
\resizebox*{0.4\textwidth}
{0.3\textheight}{\includegraphics{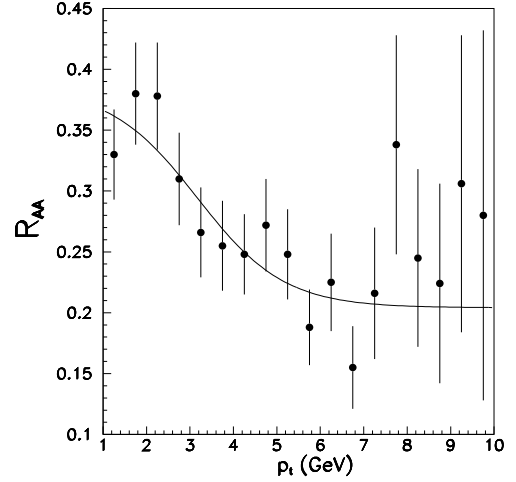}}
\par}
\caption{Nuclear suppression factor $R_{AA}$ for neutral pions from
Ref.~\cite{PHENIX2} parametrized in the interval from 1 to 10 GeV. The
parametrization is chosen as to become a constant for large $p_t$ values.} 
\label{fig4}
\end{figure}
%%%%%%%%%%%%%%%%%%%%%%%%%%%%%%%%%%%%%%%%%%%%%%%%%%%%%%%%

Gathering the elements described in Eqs.~(\ref{Wigtra})
to~(\ref{fourk}), it is straightforward to evaluate the integral
in Eq.~(\ref{distone}). Summing over all the states, the invariant
thermal distribution is given by
\be
   E\frac{d^3N}{d^3p}&=&{\mathcal {N}}\sum_{nl}
   \frac{(2l+1)}{(2\pi)}\frac{k_{nl}^2E_{nl}\ e^{-\bar{\gamma}E_{nl}/T}}
   {\sqrt{p^2+\left(\frac{\bar{\gamma}\beta
   k_{nl}}{2RT}\right)^2}}\nonumber\\
   &\times&\frac{\left|J_{l+1/2}\left(pR-i
   \frac{\bar{\gamma}\beta k_{nl}}{2T}\right)\right|^2}
   {\left[p^2-k_{nl}^2-
   \left(\frac{\bar{\gamma}\beta k_{nl}}{2RT}\right)^2\right]^2 
   + \left[\frac{\bar{\gamma}\beta p k_{nl}}{RT}\right]^2},
   \label{invdistall}
\ee
where the factor $(2l+1)$ comes from the degeneracy of a state with a
given angular momentum eigenvalue $l$ and ${\mathcal N}$
is a normalization constant.

%%%%%%%%%%%%%%%%%%%%%%%%%%%%%%%%%%%%%%%%%%%%%%%%%%%%%%%%
%\vspace{0.9cm}
\begin{figure}[t!] % fig5
%\vspace{0.4cm}
{\centering
\resizebox*{0.4\textwidth}
{0.35\textheight}{\includegraphics{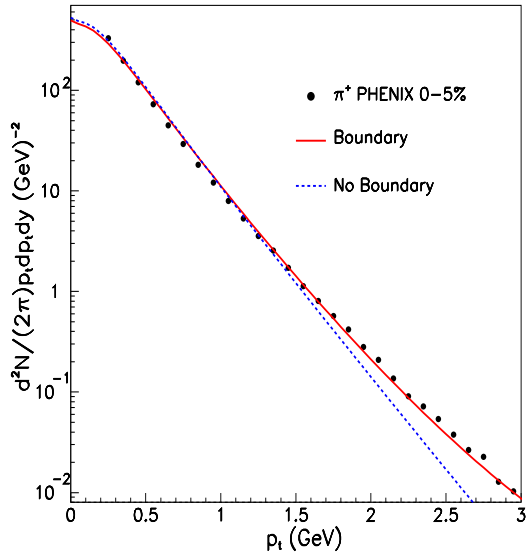}}
\par}
\caption{Invariant pion distribution as a function of $p_t$ for
$R=8$ fm, $T=120$ MeV and $\beta=0.6$ compared to data on mid rapidity
positive pions from central collisions at $\sqrt{s_{NN}}=200$ GeV
measured by PHENIX. The normalization of the model theoretical curve
has been fixed from a fit to the data.} 
\label{fig5}
\end{figure}
%%%%%%%%%%%%%%%%%%%%%%%%%%%%%%%%%%%%%%%%%%%%%%%%%%%%%%%%

Figure~\ref{fig1} shows the invariant pion distribution as a function
of $p_t$ for different combinations of the parameters $T$ and
$\beta$ and a fixed value $R=8$ fm. Figure~\ref{fig1}$a$ shows the
distribution for two different temperatures $T=100,\ 120$ MeV and
$\beta =0.5$. As expected, the effect of increasing the temperature is
to increase the inverse slope of the
distribution. Figure~\ref{fig1}$b$ shows the distribution for two   
different values of the surface expansion velocity, $\beta=0.5,\ 0.8$
and $T=100$ MeV. The effect of increasing the expansion velocity is also to
increase the inverse slope of the distribution. Figure~\ref{fig2} shows an
example of the kind of rapidity distributions obtained in the model. Notice
that the distribution is not as broad as data seem to indicate (see for
example Ref.~\cite{BRAHMS}.) This is a limitation of the spherical symmetry
assumed in the model emphasizing the need of restricting its applicability to
the description of central rapidity data.  

Figure~\ref{fig3} shows the distribution
for two different values of the confining radius $R=8,\ 10$ fm for
fixed values of $T=120$ MeV and $\beta=0.6$. The effect of increasing
the confining radius is to increase the value of $p_t$ for which
the distribution has a point of inflexion towards a concave shape. This
is to be expected since in the limit where the confining volume goes
to infinity, the value of $p_t$ for this point of inflexion should
also go to infinity as the system becomes boundless.

%%%%%%%%%%%%%%%%%%%%%%%%%%%%%%%%%%%%%%%%%%%%%%%%%%%%%%%%
%\vspace{0.9cm}
\begin{figure}[t!] % fig6
%\vspace{0.4cm}
{\centering
\resizebox*{0.4\textwidth}
{0.35\textheight}{\includegraphics{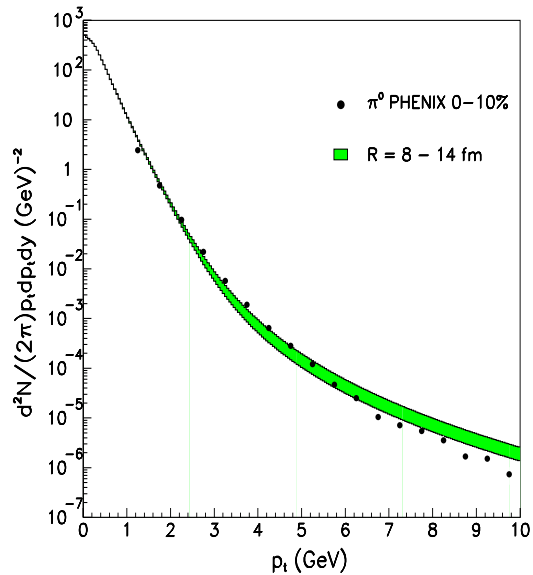}}
\par}
\caption{Invariant pion distribution as a function of $p_t$ for
$T=120$ MeV, $\beta=0.6$ for values of $R=8$ fm (upper curve) to
$R=14$ fm (lower curve), compared to data on mid rapidity
neutral pions from central collisions at $\sqrt{s_{NN}}=200$ GeV
measured by PHENIX. The normalization of the curves
has been taken from the fit to the $\pi^+$ spectrum.} 
\label{fig6}
\end{figure}
%%%%%%%%%%%%%%%%%%%%%%%%%%%%%%%%%%%%%%%%%%%%%%%%%%%%%%%%

\section{Pion spectra}\label{IV}

In order to test the model, we compare the theoretical distribution to
data on mid-rapidity pions produced in RHIC central collisions at
$\sqrt{s_{NN}}=200$ GeV~\cite{PHENIXBM,PHENIX2}. Rather than
performing an overall fit to find the optimum parameters, we fix them
to values within the commonly accepted ones to describe freeze-out
conditions. 

To include the effects of parton energy loss, we
resort to parametrize the data on $R_{AA}$ for neutral pions as
measured by PHENIX~\cite{PHENIX2}. We use the the expression
\be
   \frac{A}{e^{x-B}+1} + C\, ,
   \label{param}
\ee
and find from a fit to the data on $R_{AA}$ the values $A=0.18$,
 $B=3.17$ and $C=0.20$. This parametrization is taken in
such a way that for large $p_t$ the nuclear suppression factor remains
a constant. This is shown in Fig.~\ref{fig4}. 

First we describe data for the low $p_t$ part of the spectra. This is
done in Fig.~\ref{fig5} where we show the invariant pion distribution
as a function of $p_t$ for $R=8$ fm, $T=120$ MeV and $\beta=0.6$,
including the energy loss effect, compared to data on $\pi^+$ from
PHENIX~\cite{PHENIXBM}. The normalization of the model theoretical
curve has been calculated from a fit to the data. We notice from
Fig.~\ref{fig5} that the curve does a very good job describing the
data for all values of $p_t$. In contrast, a calculation where no
effects of a finite size are included, and thus the wave function of a
given state is simply a plane wave, does not describe the data over
the considered range when we use the same values for $T$ and $\beta$
as for the case of the calculation with a confining volume. This is
also shown in Fig.~\ref{fig5} where we notice the more steep fall-off
of the curve without finite size effects.  

Next we describe data in the range 1 GeV $< p_t <$ 10 GeV. This is
done in Fig.~\ref{fig6} where we show the invariant pion distribution
as a function of $p_t$ for $T=120$ MeV, $\beta=0.6$ and $R=8-14$ fm,
including the energy loss effect, compared to data on neutral pions
from PHENIX~\cite{PHENIX2}. The normalization of the curves has been
taken from the fit to the $\pi^+$ spectrum. The model does a good
description of data for values of the parameter $R$ between 8 and 10 fm
up to $p_t\sim 3$ GeV from where the curves start deviating from the
data that shows a steeper fall-off. This failure is however expected
since for large $p_t$ the leading particle production mechanism is the
fragmentation of fast moving partons, some of which fragment outside
the fireball region and thus are not influenced by the confining
boundary that the rest of the particles experience within the fireball.

\section{Summary and conclusions}\label{concl}

In this work we have computed the pion transverse momentum
distribution in a thermal model including transverse expansion and
accounting for the fact that the system of particles is produced within
a small volume up to freeze-out, in the context of relativistic
heavy-ion collisions. The finite size has been introduced by requiring
that the system's wave functions vanish at the boundary but remain
finite at the origin. The net
effect is a broadening of the particle transverse momentum
distribution with respect to a simple thermal model without a
boundary. The physical origin of this broadening is the Heisenberg
uncertainty principle since as the states are more localized in
coordinate space, their spread is wider in momentum space. Energy loss
has also been considered by a parametrization of the experimentally
measured nuclear suppression factor $R_{AA}$ where we require that
this becomes constant for large $p_t$ values.

The model does a very good job describing data for low to intermediate
$p_t$ in the range $0<p_t\lesssim3$ GeV for commonly accepted values of the
temperature $T=120$ MeV and the surface transverse expansion velocity
$\beta = 0.6$ (corresponding to an average transverse expansion
velocity $\langle\beta\rangle\simeq 0.5$) when the system's 
radius $R$ at freeze-out is in the range 8 fm $\lesssim R \lesssim$ 10
fm. This is best seen when comparing the model to data on charged
pions produced in central Au + Au collisions. The description of data
for $p_t$ in the range $3 \lesssim p_t < 10$ is not as good as can be
seen from a comparison of the model to the spectrum of neutral pions
produced in central Au + Au collisions. This behavior is not
surprising though, since for large $p_t$ it is well known that the
leading particle production mechanism is the fragmentation of fast
moving partons, some of which fragment outside the fireball region and
thus are not influenced by the confining boundary, or mean field, that
the rest of the particles experience within the fireball.

We thus conclude that the thermal features of the spectrum, which are
obtained from the low $p_t$ region, are best highlighted when
accounting for the finite size of the system at decoupling. It is of
course interesting to check the consistency of this treatment when
extended to the case of fermions, and in particular protons. Finally, it is
worth to note that finite size effects should also play a role
during the partonic phase of the collision~\cite{Wong} and in particular that
they can also influence the recombination scenario. This kind of analyses is
for the future. 

\section*{Acknowledgments}

The authors thank G. Paic for his valuable comments and
suggestions and I. Dominguez for having prepared a PYTHIA simulation
for comparison with our work. A.A. and L.M.M. thank the staff at CBPF
for their help during a summer visit when part of this work was
done. Support for this work has been received in part by PAPIIT-UNAM
under grant number IN107105  and CONACyT under grant numbers 40025-F
and bilateral agreement CONACyT-CNPq J200.556/2004 and 491227/2004-3.

\end{document}